\documentclass[12pt]{article}
\usepackage[numbers,sort&compress]{natbib}
\usepackage[utf8]{inputenc}
\usepackage[margin=1.0in]{geometry}
\usepackage{amscd,amssymb,amsxtra}
\usepackage{amsmath, amsthm, amssymb}
\usepackage{physics}
\usepackage[colorlinks=true, linktocpage=true, linkcolor=blue]{hyperref}
\usepackage{authblk}

\numberwithin{equation}{section}

\usepackage{color}
\usepackage{tikz-cd}
\usepackage{comment}
\usepackage{tikz}
\usepackage{subfig}

\usepackage{setspace}
\usepackage{lineno}

\begin{document}


\title{Mass Spectra and Decay of Mesons under Strong External Magnetic Field }

\author{Shuyun Yang\thanks{yangsy@mails.ccnu.edu.cn}}
\author{ Meng Jin\thanks{jinmeng@mail.ccnu.edu.cn}}
\author{Defu Hou\thanks{Corresponding author: houdf@mail.ccnu.edu.cn}}
\affil[1]{\footnotesize Institute of Particle Physics and Key Laboratory of Quark and Lepton Physics (MOS), Central China Normal University,
Wuhan 430079, China}
\maketitle

\begin{abstract}
  We study the mass spectra and decay process of $\sigma$ and $\pi_0$ mesons under strong external magnetic field. To achieve this goal, we deduce the thermodynamic potential in a two-flavor, hot and magnetized Nambu-Jona-Lasinio model. We calculate the energy gap equation through the random phase approximation (RPA). Then we use Ritus method to calculate the decay triangle diagram and self-energy in the presence of a constant magnetic field B. Our results indicate that the magnetic field has little influence on the mass of $\pi_0$ at low temperatures. While for quarks and $\sigma$ mesons, their mass changes obviously, which reflects the influence of magnetic catalysis (MC). The presence of magnetic field accelerates the decay of the meson while the presence of chemical potential will decrease the decay process.
\end{abstract}

\maketitle
\section{Introduction}

  Over the years, studies on the nature and state of strong interacting substances under extreme conditions have attracted much attention, where extreme conditions include high temperature and finite baryon chemical potential~\cite{Goldberger:1958tr, GellMann:1968rz, Soloveva:2019xph}, as well as strong magnetic fields. In this paper we discuss the effects of finite magnetic fields on strong interacting matter~\cite{Fayazbakhsh:2012vr,Fayazbakhsh:2013cha,Fayazbakhsh:2013em,Kayanikhoo:2019ugo,Gong:2019khm,Xu:2021,Xu:2020,Andersen:2014xxa,Miransky:2015ava,Li:2016tel,Siddique:2021smf}. Magnetic field changes are closely related to high-energy nuclear collisions, dense stars and cosmic phase transitions. The maximum magnetic field observed in nature is about $10^{12}-10^{13}$ $\mathrm{Gauss}$ in pulsars, the maximum magnetic field on the surface of some magnetospheric stars is around $10^{14}-10^{15}$ $\mathrm{Gauss}$, and its internal field is estimated to be $10^{18}-10^{20}$ $\mathrm{Gauss}$. In the early stage of the RHIC non-central heavy ion collision, there is also an evidence of a very strong and instantaneous magnetic field. Depending on the energy of the collision and impact parameters, magnetic field produced in RHIC is about $eB\sim1.5m_{\pi}^{2}\sim0.03$ $\mathrm{GeV^{2}}$, and $eB\sim15 m_{\pi}^{2}\sim0.3$ $\mathrm{GeV^{2}}$ in LHC~\cite{Ferrer:2010wz,Selyuzhenkov:2005xa,Kharzeev:2007jp,Skokov:2009qp}. On the other hand, the quark-gluon plasma  produced in high energy heavy-ion collisions, went through many stages in the process of evolution. A large number of hadrons including $\pi$ are produced, freeze out and then survive in the final state~\cite{Ayala:2002qy}. Therefore, the existence of the background magnetic field generated in the heavy ion experiment may affect the properties of the early "charged quarks" in the collision. Although this strong magnetic field lasts a very short time and disappears very fast, it may affect the properties of the hadrons formed by these "magnetized" quarks. Even the properties of neutral mesons may be affected by external magnetic fields produced in the early stage of heavy ion collisions~\cite{Kharzeev:2007jp,Skokov:2009qp}
 
This paper is based on the two flavor NJL model~\cite{Nambu:1961fr,Klevansky:1992qe,Hatsuda:1994pi,Buballa:2003qv}. We calculate the energy gap equation by the mean field approximation. And we use Ritus method to calculate the decay triangle diagram and self-energy in the magnetic field through Random Phase Approximation (RPA), then we obtained the meson decay width, meson-quark coupling constant. Our result shows that as quarks which constitutes mesons are magnetized under magnetic field, the mass of $\sigma$ grows whereas the mass of $\pi_0$ changes very little at low temperature, the coupling constant $g_{\sigma qq}$ and $g_{\pi_{0}qq}$ are significantly larger than that without magnetic field.

 The organization of the paper is as follows. In Sec.~\ref{sec:E}, we give the effective thermodynamic potential . In Sec.~\ref{sec:Decay}, we calculate the mass spectra of mesons and the decay constant of $\sigma\rightarrow\pi_{0}\pi_{0}$. Sec.~\ref{sec:N} is the numerical results of our work. Finally we summarize and discuss some possible extension of our work in Sec.~\ref{sec:S}.

\section{Effective thermodynamic potential}\label{sec:E}
  The Lagrangian density of SU(2) NJL model is
\begin{equation}
\mathcal{L}_{NJL}=\bar\psi(i\gamma^\mu\partial_\mu-m_0)\psi+G[(\bar\psi\psi)^2+(\bar\psi{i}\gamma_5\boldsymbol{\tau}\psi)^2].
\end{equation}

The Lagrangian density has $SU_V(2)\times{SU_A(2)}\times{U_V(1)}$ symmetry, where $SU_V(2)$ corresponding to the chiral symmetry, $SU_A(2)$ corresponding to the conservation of isotopic spin, $U_V(1)$ corresponding to the conservation of baryon Numbers. In the formula, $\psi$ and $\bar\psi$ are quark fields. When the isospin symmetry is satisfied, we have $m_{\mu}=m_{d}=m_{0}$. $D^{\mu}=\partial^{\mu}+ieA^{\mu}$ is covariant differentiation, $A^{\mu}=\delta_{0}^{\mu}A^{0}$ and $A^{0}=-iA^{4}$ are gauge fields. G is the four quark coupling constant. The pauli matrix $\tau_{i}(i=1,2,3)$ is defined in the isospin space.

The thermodynamic potential can be expressed as
\begin{equation}
\Omega=-\frac{T}{V}lnZ.
   \end{equation}
with the partition function
\begin{equation}
 Z(T,\mu,V)=\int[d\bar{\psi}][d\psi]e^{\intop_{0}^{\beta}d\tau\int d^{3}\boldsymbol{x}(\mathcal{L}+\bar{\psi}\mu\gamma_{0}\psi)}.
   \end{equation}
where V is the volume, $\beta=\frac{1}{T}$. $\mu=dig(\mu_{u},\mu_{d})$ is the chemical potential of quark. Following the calculations such as~\cite{kapusta_gale_2006,KUNIHIRO1988385}, we have the thermodynamic potential under the mean field approximation,
\begin{equation}
\Omega=-\frac{T}{V}lnZ=-\frac{T}{V}lnTre^{-\beta\int d^{3}\boldsymbol{x}
 (\mathcal{L}+\bar{\psi}\mu\gamma_{0}\psi)}=\Omega_{q}+\frac{(m_{q}-m_{0})^{2}}{4G}+const.
   \end{equation}
where $\Omega_{q}$ is the contribution of the quark part
\begin{equation}
\Omega_{q}(T,\mu)=-T\mathop{\sum}_{n}\int\frac{d^{3}\boldsymbol{p}
 }{(2\pi)^{3}}Trln(\beta S^{-1}(i\omega_{n},\boldsymbol{p})).
   \end{equation}
In this formula $S^{-1}(p)=\gamma^{\mu}p_{\mu}-m_{q},
p^{0}=i\omega_{n}=(2n+1)\pi T$. It is known that
\begin{equation}
 Trln(\gamma^{\mu}p_{\mu}-m_{q})=lnDet(\gamma^{\mu}p_{\mu}-m_{q})=2N_{c}N_{f}ln(p^{2}-m_{q}^{2}).
    \end{equation}
Using of the summation formula~\cite{Nambu:1961tp}
 \begin{equation}
  T\mathop{\sum}_{n}ln(\beta^{2}(\omega_{n}^{2}+\lambda_{k}^{2}))=\lambda_{k}+2Tln(1+e^{-\beta\lambda_{k}})  ,
    \end{equation}
 one can get
 \begin{equation}
\Omega_{q}(T,\mu)=-2N_{c}N_{f}\int\int\frac{d^{3}\boldsymbol{p}
 }{(2\pi)^{3}}\{E_{p}+Tln(1+e^{-\beta(E_{p}-\mu)}+Tln(1+e^{-\beta(E_{p}+\mu)})\}.
    \end{equation}
where, $E_{p}=\sqrt{p_{3}^{2}+m_{q}^{2}}$. Regardless of the chemical potential, $\mu=0$, the formula will change to
   \begin{equation}
\Omega_{q}=-2N_{c}N_{f}\int\frac{d^{3}\boldsymbol{p}
 }{(2\pi)^{3}}\{E_{p}+2Tln(1+e^{-\beta E_{p}})\}.
    \end{equation}
If the magnetic field is added, one will find that
\begin{equation}
2 N_{f} \int \frac{d^{3}\boldsymbol{p}
 }{(2 \pi)^{3}} \rightarrow \sum_{f, n} \alpha_{n} \frac{\left|Q_{f} B\right|}{2 \pi} \int \frac{d p_{3}}{2 \pi},
\end{equation}
  then, the thermodynamic potential becomes
\begin{equation}
\begin{aligned}
\label{1}
\Omega_{q}(T,\mu)=&-N_{c}\sum_{f,n}\alpha_{n}\frac{\left|Q_{f}B\right|}{2\pi}\int\frac{dp_{3}}{2\pi}\{E_{f}+T\ln\left(1+e^{-\beta(E_{f}-\mu)}\right)+T\ln\left(1+e^{-\beta(E_{f}+\mu)}\right)\}.
\end{aligned}
\end{equation}
 where, $E_f=\sqrt{p_{3}^{2}+2 n\left|Q_{f} B\right|+m_{q}^{2}}$, $\alpha_{n}=2-\delta_{n0}$, charge $Q=diag(Q_{u},Q_{d})$ to the external magnetic field $\textbf{B}=(0,0,B)$ in z-direction, $f$ is flavor $u$ or $d$.

\section{Decay process in two flavor NJL model}\label{sec:Decay}
  The energy gap equation can be derived from the derivative of the thermodynamic potential. From energy gap equation we can calculate the mass of quark and the single loop self-energy through RPA~\cite{Klevansky:1992qe,Hatsuda:1994pi,Buballa:2003qv,Zhuang:1994dw} in a two flavor NJL model. As for the selection of magnetical quark propagators, usually we have two methods: Ritus scheme~\cite{Ritus:1972ky,Ritus:1978cj,Leung:2005yq,Elizalde:2000vz,Menezes:2008qt,Ferrer:2010wz,Fukushima:2009ft,Mao:2016lsr,Mao:2016,Mao:2019avr} and Schwinger scheme~\cite{Andersen:2014xxa,Miransky:2015ava}. In this paper, the Ritus scheme is used to deal with propagators, which is described in coordinate space,
\begin{equation}
\label{3}
S_{f}(x,y)=\mathop{\sum}_{n}\int\frac{d^{3}\tilde{p}}{(2\pi)^{3}}e^{-i\tilde{p}(x-y)P_{n}(x_{1},p_{2})D_{f}(\bar{p})P_{n}(y_{1},p_{2})},
  \end{equation}
where,
\begin{equation}
D_{f}(\bar{p})=\gamma\cdot\bar{p}-m_{q},
 \end{equation}
and the magnetic field related terms are
\begin{equation}
P_n(x_1,p_2)=\frac{1}{2}[g_n^{sf}(x_1,p_2)+I_ng_{n-1}^{sf}(x_1,p_2)]+\frac{iS_f}{2}[g_n^{sf}(x_1,p_2)-I_ng_{n-1}^{sf}(x_1,p_2)]\gamma_1\gamma_2.
  \end{equation}
the term $g_n^{sf}(x_1,p_2)=\phi_n(x_1-s_fp_2/|Q_fB|)$ is determined by hermite polynomial $H_{n}(\zeta)$.
\begin{equation}
\phi_n(\zeta)=(2^nn!\sqrt{\pi}|Q_fB|^{-1/2})^{-\frac{1}{2}}e^{-\frac{\zeta^2|Q_fB|}{2}}H_n(\zeta/|Q_fB|^{-1/2}).
  \end{equation}
  Here, $\tilde{p}=(p_{0},0,p_{2},p_{3})$ is the momentum of the Fourier transform, $\bar{p}=(p_{0},0,-s_{f}\sqrt{2n|Q_{f}B|},p_{3})$ is conserved Ritus momentum. $s_{f}=sgn(Q_{f}B)$, $Q_{f}=(2/3,-1/3)$, f means quark flavor. $I_{n}=1-\delta_{n0}$, n is the landau energy level. 

\subsection{Mass spectra of mesons}\label{sec:M}
  To illustrate the Ritus scenario clearly, we use neutral mesons as a simple example. In this case, the meson momentum $k=(\omega,k_{1},k_{2},k_{3})$ is conserved for neutral mesons which do not interact with the magnetic field. The corresponding meson polarization functions in the momentum space are the Fourier transforms of their expressions in the coordinate space,
 \begin{equation}
\begin{aligned}
&\mathcal{D}_M(k)=\int d^4(x-y)e^{ik(x-y)}\mathcal{D}_M(x,y),\\
&\Pi_M(k)=\int d^4(x-y)e^{ik(x-y)}\Pi_M(x,y).
\end{aligned}
  \end{equation}
Using random phase approximation , one can get
  \begin{equation}
\mathcal{D}_M(k)=\frac{2G}{1-2G\Pi_M(k)}.
  \end{equation}
The mass of a meson is defined by the pole of the denominator at zero momentum
 \begin{equation}
\label{2}
  1-2G\Pi_M(\omega=m_M,\boldsymbol{0})=0.
  \end{equation}
 The polarization function is~\cite{Mao:2018dqe},
   \begin{equation}
\Pi_M(\omega,\boldsymbol{0})=J_1-(\omega^2-\epsilon^2_M)J_2(\omega^2).
  \end{equation}
   where,
  \begin{equation}
   J_1=2N_cN_f\int\frac{d^3\boldsymbol{p}}{(2\pi)^3}\frac{tanh(\frac{E_f}{2T})}{E_f},
  \end{equation}
   and
   \begin{equation}
  J_2(\omega^2)=-2N_cN_f\int\frac{d^3\boldsymbol{p}}{(2\pi)^3}\frac{tanh(\frac{E_f}{2T})}{E_f(4E^2_f-w^2)}.
 \end{equation}
 with $\epsilon_{\pi_0}=0$ and $\epsilon_{\sigma}=2m_q$.

 Under zero magnetic field
   \begin{equation}
   \Pi_M(\omega,\boldsymbol{0})=2N_cN_f\int\frac{d^3\boldsymbol{p}}{(2\pi)^2}\frac{E_f-\epsilon^2_M}{E_f^2-\frac{w^2}{4}}(1-2f_{F}(E_f)).
    \end{equation}
where $E_f=\sqrt{{\boldsymbol{p}}^{2}+m_{q}^{2}}$ for the distribution function $f_{F}(E_{f})=\frac{1}{e^{\beta E_{f}+1}}$,
   After considering the magnetic field, one can make the substitution in Eq.~(\ref{1}). Therefore, the self-energy of mesons under external magnetic field can be further obtained,
\begin{equation}
\begin{aligned}
\Pi_{M}^{'}(\omega,\boldsymbol{0})=N_{c}\sum_{f,n}\alpha_{n}\frac{\left|Q_{f}B\right|}{2\pi}\int\frac{dp_{3}}{2\pi}\frac{1-f_{F}\left(E_{f}-\mu\right)-f_{F}\left(E_{f}+\mu\right)}{E_{f}}\\
+(\omega^{2}-\epsilon_{M}^{2})N_{c}\sum_{f,n}\alpha_{n}\frac{\left|Q_{f}B\right|}{2\pi}\int\frac{dp_{3}}{2\pi}\frac{1-f_{F}\left(E_{f}-\mu\right)-f_{F}\left(E_{f}+\mu\right)}{E_{f}\left(4E_{f}^{2}-w^{2}\right)}  ,
 \end{aligned}
\end{equation}
  
\subsection{The decay rate for $\sigma\rightarrow\pi_{0}\pi_{0}$}
  Elementary particles have tendency to decay. For a given interaction, the larger the mass difference between the initial particles and the decay products, the faster the decay proceeds.
\begin{figure}[ht]
  \centering
\includegraphics[width=3.12in,height=1.39in]{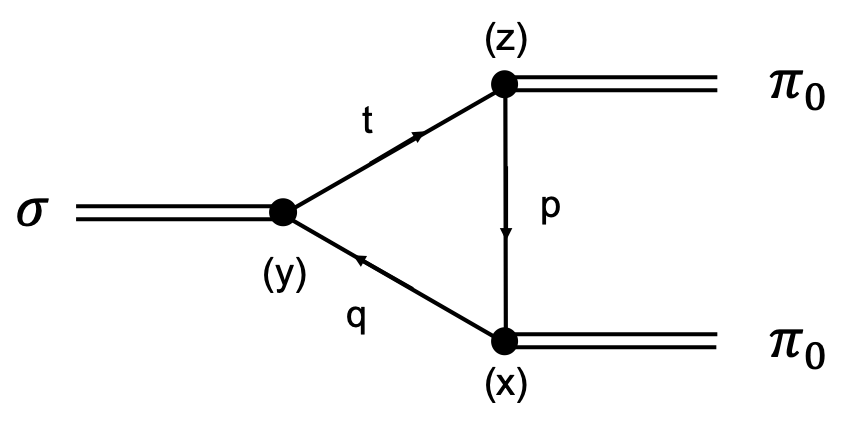}\\
  \caption{Feynman diagram for the process $\sigma\rightarrow\pi_{0}\pi_{0}$}
\end{figure}
 Here we define the decay rate of a particle by a Lorentz conservation matrix M~\cite{Zhu:2001iu}
 \begin{equation}
\frac{d\Gamma_{\sigma\rightarrow2\pi_{0}}}{d\Omega}=\frac{1}{32\pi^{2}}\frac{|P|}{m_{\sigma}^{2}}|M(T,\mu)|^{2},
    \end{equation}
    where $\pi_{0}$ momentum is $|P|=\sqrt{\mbox{\ensuremath{\frac{m_{\sigma}^{2}}{4}-m_{\pi_{0}}^{2}}}} $ and $M(T,\mu)=g_{\sigma}g_{\pi_{0}}^{2}A_{\sigma\pi_{0}\pi_{0}}$.

At a finite temperature and density, the decay rate of $\sigma\rightarrow\pi\pi$ process is~\cite{Zhuang:2000tz}
 \begin{equation}
   \begin{aligned}\Gamma_{\sigma\rightarrow2\pi_{0}}(T,\mu) & =\Gamma_{\sigma\rightarrow2\pi_{0}}(T,\mu)\\
 & =\frac{3}{8\pi}\frac{\sqrt{m_{\sigma}^{2}/4-m_{\pi_{0}}^{2}}}{m_{\sigma}^{2}}g_{\sigma}^{2}g_{\pi_{0}}^{4}\left|A_{\sigma\pi_{0}\pi_{0}}(T,\mu)\right|^{2}\left[1+2f_{B}\left(\frac{m_{\sigma}}{2}\right)\right]
\end{aligned}.   
 \end{equation}
where, $f_{B}\left(\frac{m_{\sigma}}{2}\right)=\frac{1}{e^{\frac{m_{\sigma}}{2}}-1}$ is the Bose-Einstein distribution function.
 
   $\sigma$ coupling constant and $\pi$ coupling constant are respectively defined by
 \begin{equation}
 g_{\sigma qq}^{-2}=\frac{\partial\Pi_{\sigma}(k_{0},\boldsymbol{0})}{\partial k_{0}^{2}};\\
 g_{\pi_{0} qq}^{-2}=\frac{\partial\Pi_{\pi_{0}}(k_{0},\boldsymbol{0})}{\partial k_{0}^{2}} .
    \end{equation}
The triangle factor $A_{\sigma\pi_{0}\pi_{0}}$ is defined by
 \begin{equation}
 i A_{\sigma \pi_{0} \pi_{0}}(T, \mu)=-T_{r} \int \frac{d^{4} q}{(2 \pi)^{4}}\left[g_{\sigma qq} \Gamma_{\sigma} i S_{u}(x, y) g_{\pi_{0} qq} \Gamma_{\pi} i S_{u}(y, z) g_{\pi_{0} qq} \Gamma_{\pi} i S_{u}(z, x)\right]  ,
  \end{equation}
   the vertex of the meson is expressed as
\begin{equation}
\Gamma_{M}=\left\{\begin{array}{cc}{1} & {M=\sigma}  \\ {i \tau_{3} \gamma_{5}} & {M=\pi_{0}}\end{array}\right.
\end{equation}
In Eq.~(\ref{3}), we have defined the propagator of meson in coordinate space. Here, we let
\begin{equation}
\begin{aligned} P_{n}\left(x_{1}, q_{2}\right)&=\frac{1}{2}\left[g_{n}^{S_{f}}\left(x_{1}, q_{2}\right)+I_{n} g_{n-1}^{S_{f}}\left(x_{1}, q_{2}\right)\right]+\frac{i S_{f}}{2}\left[g_{n}^{S_{f}}\left(x_{1}, q_{2}\right)-I_{n} g_{n-1}^{S_{f}}\left(x_{1}, q_{2}\right)\right] \gamma_{1} \gamma_{2} \\
&=\frac{1}{2}\left[f_{k+}(x)+f_{k-}(x)\right]+\frac{i S_{f}}{2}\left[f_{k+}(x)-f_{k-}(x)\right] \gamma_{1} \gamma_{2} \\
&=A_{x}+i S_{f} B_{x} \gamma_{1} \gamma_{2} \end{aligned}.
\end{equation}
Then, we can get the triangle factor $A_{\sigma\pi_{0}\pi_{0}}$
\begin{equation}
\begin{aligned}
A_{\sigma \pi_{0} \pi_{0}}(x, y, z)=& 2 i^{6} N_{c} g_{\sigma qq}g_{\pi_{0} qq}^{2} \operatorname{Tr}\left[1 \cdot S_{u}(x, y) \cdot r_{5} \cdot S_{u}(y, z) \cdot r_{5} \cdot S_{u}(z, x)\right] \\=& 2 i^{2} N_{c} g_{\sigma qq}g_{\pi_0 qq}^{2}\sum_{ n, n^{\prime}, n^{\prime \prime}} \int \frac{d^{3} \tilde{p} d^{3} \tilde{q}^{3} \tilde{q}}{(2 \pi)^{9}} e^{-i \tilde{q}(x-y)-i \tilde{l}(y-z)-i \tilde{p}(z-x)} \\
& \cdot \operatorname{Tr}\left[P_{n}\left(x_{1}, q_{2}\right) \frac{r \cdot \bar{q}+m_{q}}{\bar{q}^{2}-m_{q}^{2}} P_{n}\left(y, q_{2}\right) \cdot P_{n^{\prime}}\left(y_{1}, t_{2}\right) \frac{-r \cdot \bar{l}+m_{q}}{\bar{q}^{2}-m_{q}^{2}} P_{n^{\prime}}\left(z_{1}, t_{2}\right)\right.\\
&\left.\cdot P_{n^{\prime \prime}}\left(z_{1}, p_{2}\right) \frac{r \cdot \bar{p}+m_{q}}{\bar{q}^{2}-m_{q}^{2}} P_{n}^{\prime \prime}\left(x_{1}, p_{2}\right)\right] \end{aligned}.
\end{equation}
Normalization was carried out by the following formula~\cite{Fukushima:2009ft}
\begin{equation}
\left\{\begin{array}{ll}{\int d x f_{k+}(x) f_{l+}(x)=\delta_{k, l}} & {\Longrightarrow f_{k+} f_{k+} \quad \text {remain}} \\ {\int d x f_{k+}(x) f_{l-}(x)=\delta_{k, l-1}(l \geq 1)} & {\Longrightarrow f_{k+} f_{k-}=0} \\ {\int d x f_{k-}(x) f_{l-}(x)=\delta_{k, l}} & {\Longrightarrow f_{k-} f_{k-} \quad \text {ramain}} \\ {\int d x f_{k-}(x) f_{l+}(x)=\delta_{k-1, l}(k \geq 1)} & {\Longrightarrow f_{k-} f_{k+}=0}\end{array}\right.
\end{equation}
and $e^{-i\widetilde{q}(x-y)-i\widetilde{l}(y-z)-i\widetilde{p}(z-x)}$ in coordinate space can be changed to $(\bar{k},\bar{p}_{3})$ momentum space by Fourier transform. For $y-z\neq0$ and $z-x\neq0$, the relationship of momentum conservation becomes
\begin{equation}
\begin{aligned}
\widetilde{l}=\widetilde{q}+\widetilde{k}  \\
\widetilde{p}=\widetilde{q}+\widetilde{p}_{3}
\end{aligned},
  \end{equation}
 Finally, we have the triangle factor
  \begin{equation}
\begin{aligned}
A_{\sigma \pi_{0} \pi_{0}}\left(k, p_{3}\right)=& 2 m g_{\sigma qq}g_{\pi_{0} qq}^{2} N_{c} \sum_{f, n} \alpha_{n} \frac{\left|Q_{f} B\right|}{2 \pi} \int \frac{d q_{3}}{2 \pi} \frac{f_{F}\left(E_{q}-\mu \right)-f_{F}\left(-E_{q}-\mu \right)}{2 E_{q}} \\ & \frac{\frac{S^{2}}{2}-2 S E_{q}^{2}-\left(4 m_{\pi}^{2}+2 S\right) \boldsymbol{q} \cdot \boldsymbol{p_{3}}+8(\boldsymbol{q} \cdot \boldsymbol{p_{3}})^{2}}{\left(S-4 E_{q}^{2}\right)\left[\left(m_{\pi}^{2}-2 \boldsymbol{q} \cdot \boldsymbol{p_{3}}\right)^{2}-S E_{q}^{2}\right]}
\end{aligned}.
 \end{equation}
where $S=m_{\sigma}^2$,$|\boldsymbol{p_{3}}|=\sqrt{m_{\sigma}^{2}/4-m_{\pi_{0}}^{2}}$,$|\boldsymbol{q}|=\sqrt{q_{3}^{2}+2n|Q_{f}eB|}$. Since we only consider the decay process of $\sigma\rightarrow\pi_{0}\pi_{0}$, the decay constant can be approximated as
  \begin{equation}
\begin{aligned}\Gamma_{\sigma\rightarrow\pi_{0}\pi_{0}}(T,\mu) & =\frac{3}{8\pi}\frac{\sqrt{m_{\sigma}^{2}/4-m_{\pi_{0}}^{2}}}{m_{\sigma}^{2}}g_{\sigma qq}^{2}g_{\pi_{0}qq}^{4}\left|A_{\sigma\pi_{0}\pi_{0}}(T,\mu)\right|^{2}\left[1+2f_{B}\left(\frac{m_{\sigma}}{2}\right)\right]\end{aligned}
 \end{equation}

\section{Numerical Results}\label{sec:N}

\begin{figure}[htb]
\hspace{-.05\textwidth}
\begin{minipage}[t]{.4\textwidth}
\includegraphics*[width=\textwidth]{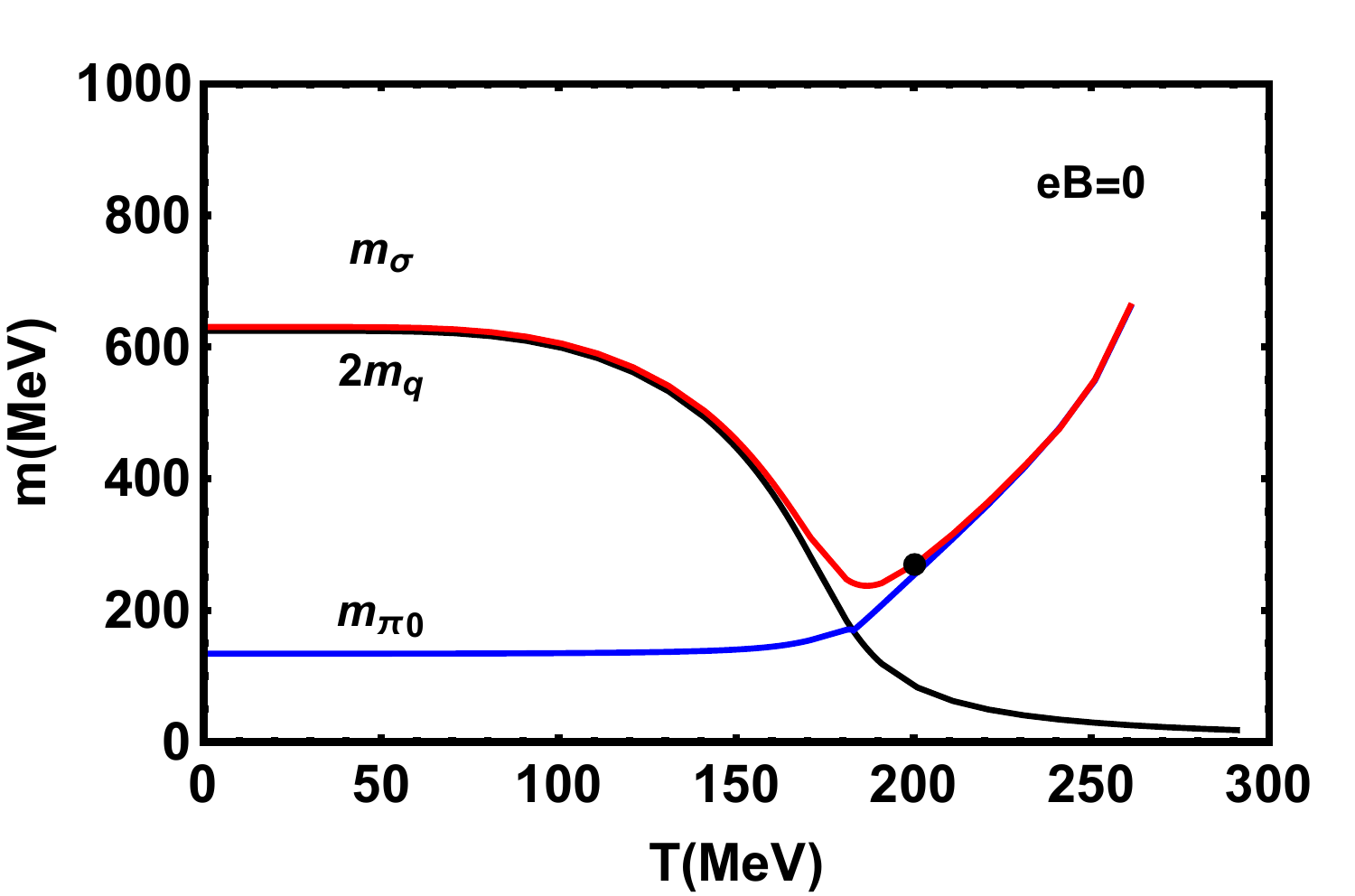}\\
\centerline{(a)}
\end{minipage}
\hspace{.05\textwidth}
\begin{minipage}[t]{.4\textwidth}
\hspace{-.05\textwidth} \scalebox{.84}
{\includegraphics*[width=1.17\textwidth]{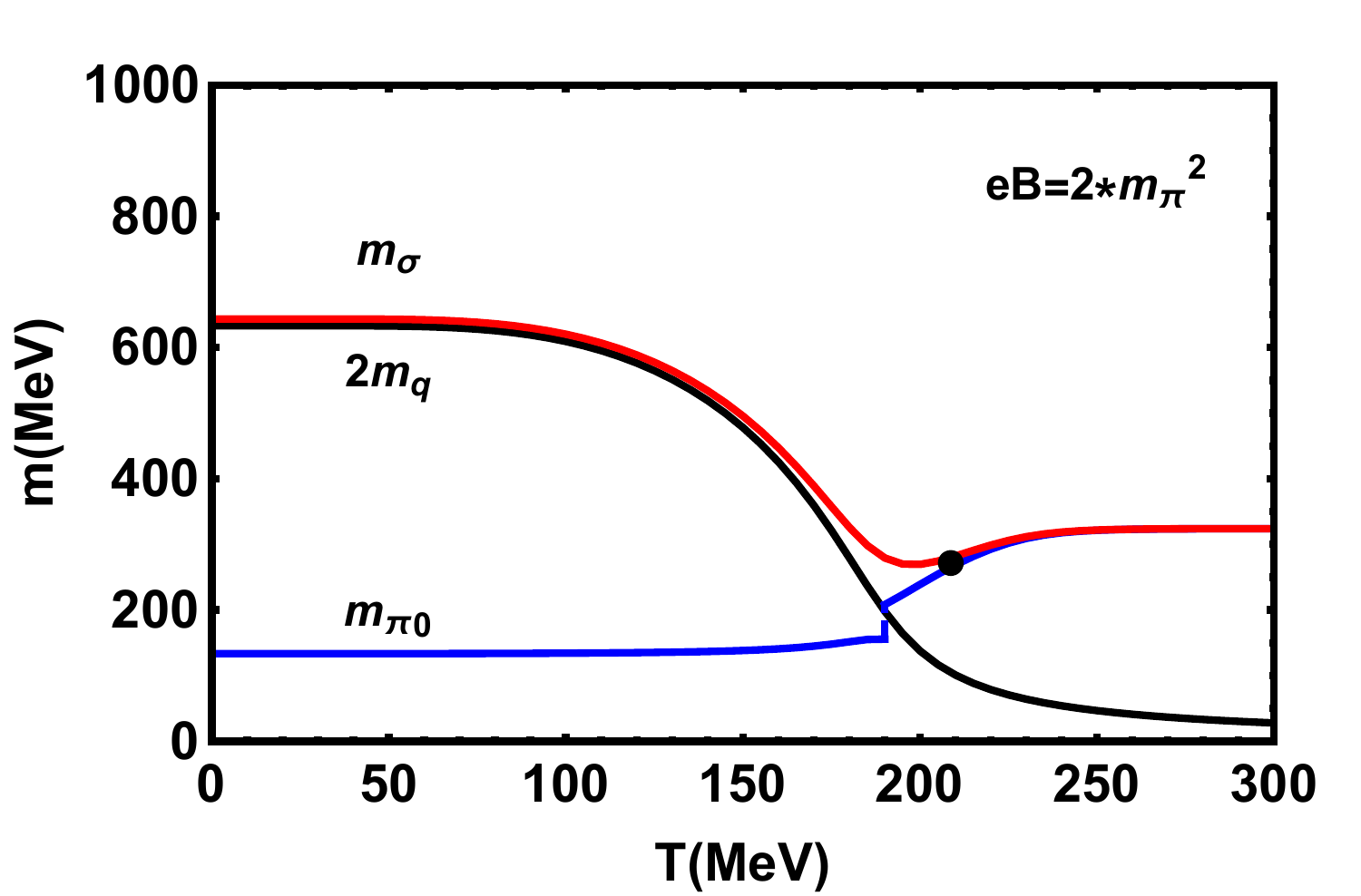}}\\
\centerline{(b)}
\end{minipage}
\centering \hspace{-.05\textwidth}
\begin{minipage}[t]{.4\textwidth}
\includegraphics*[width=\textwidth]{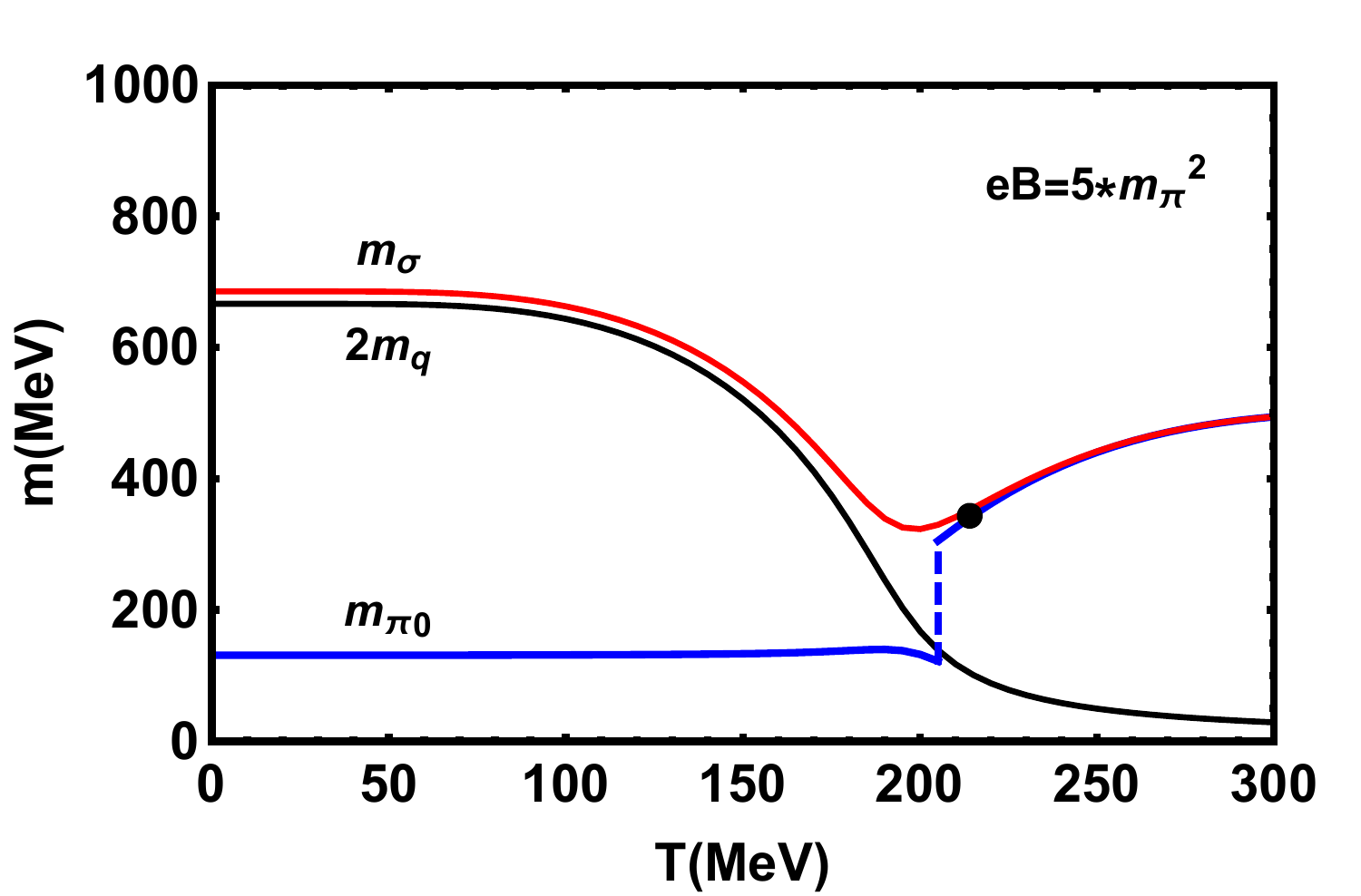}\\
\centerline{(c)}
\end{minipage}
\hspace{.05\textwidth}
\begin{minipage}[t]{.4\textwidth}
\hspace{-.05\textwidth} \scalebox{.84}
{\includegraphics*[width=1.17\textwidth]{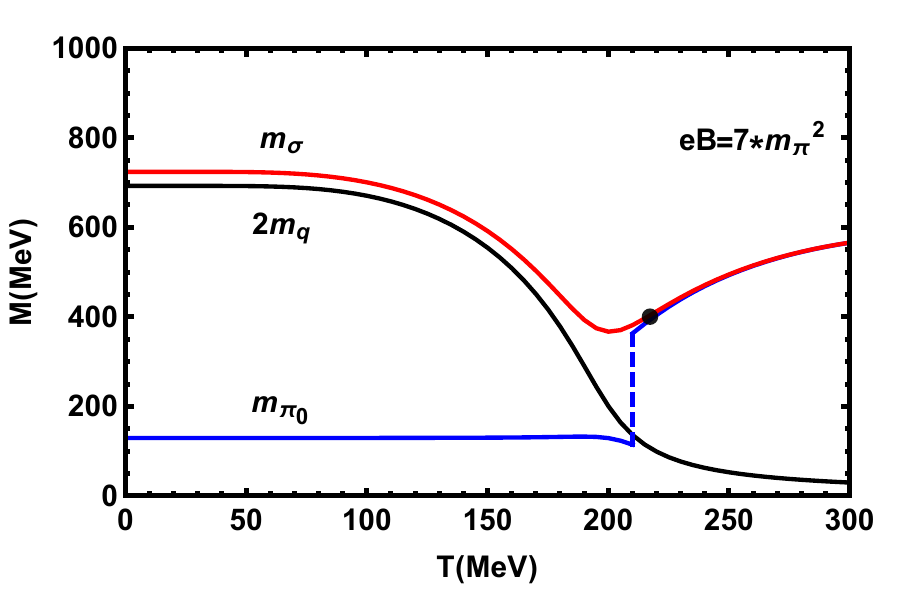}}\\
\centerline{(d)}
\end{minipage}
\parbox{15cm}
{\caption{The masses of quark, $\sigma$ and $\pi_0$ mesons at  $ eB =0, 2, 5, 7 m_{\pi}^{2}$ when $\mu=0$.}
 \label{fig2}}
\end{figure}

The chiral condensate or the dynamical quark mass is controlled by the minimum of the thermodynamic potential~\cite{Gatto:2012sp,Preis:2012fh,Andersen:2014xxa,Miransky:2015ava,Mao:2019avr}. Using the thermodynamic potential from Eq.~(\ref{1}), we have
\begin{equation}
\frac{\partial\varOmega_{mf}}{\partial m_{q}}=0.
 \end{equation}

In this work, our choice of parameters are $m_{0}$ = 5 MeV, $\Lambda$ = 653.3 MeV, $G= 4.93*10^{-6}$  $\mathrm{MeV^{2}}$. Hard-cutoff regularization scheme is adopted to deal with the integral in our work, there are other kinds of regularization schemes such as soft-cutoff regularization scheme and pauli-villars regularization scheme, one may find these methods in~Ref. \cite{Mao:2016fha}.

The diagrams of mass variation corresponding to the temperature under different external magnetic fields $ eB =0, 2, 5, 7 m_{\pi}^{2}$ are plotted in Fig.~\ref{fig2}. We observe that when the temperature increases, the mass of the quark decreases first slowly then sharply, and finally changes slowly after $ T = 200$ MeV. For a fixed temperature, the quark mass gradually increase with the increase of magnetic field.

\begin{figure}[h]
\hspace{-.05\textwidth}
\begin{minipage}[t]{.4\textwidth}
\includegraphics*[width=\textwidth]{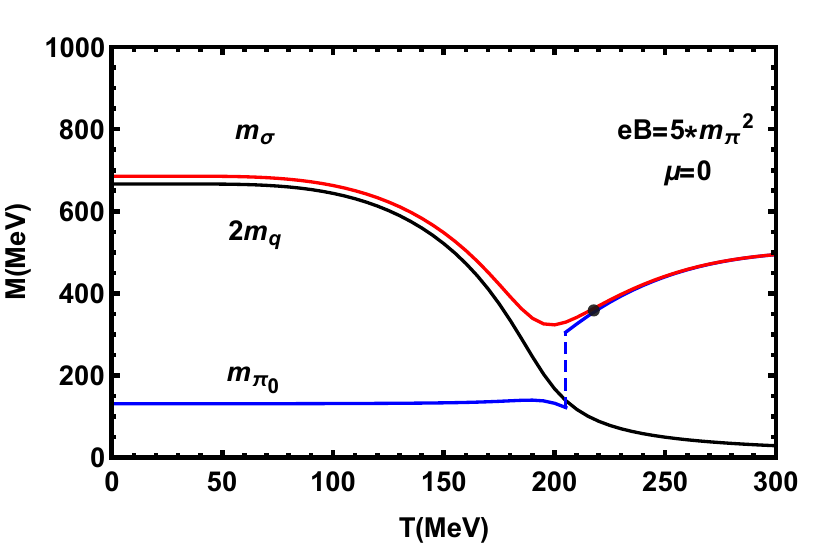}\\
\centerline{(a)}
\end{minipage}
\hspace{.05\textwidth}
\begin{minipage}[t]{.4\textwidth}
\hspace{-.05\textwidth} \scalebox{.84}
{\includegraphics*[width=1.17\textwidth]{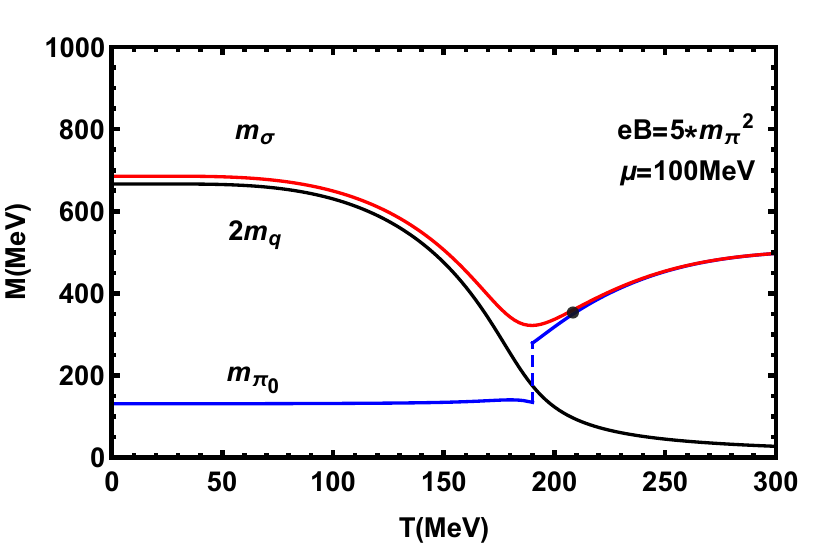}}\\
\centerline{(b)}
\end{minipage}
\centering \hspace{-.05\textwidth}
\begin{minipage}[t]{.4\textwidth}
\includegraphics*[width=\textwidth]{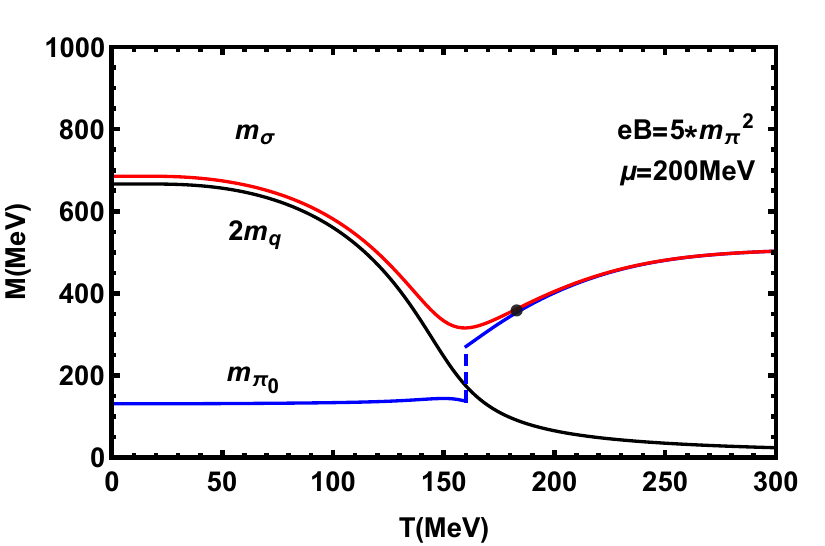}\\
\centerline{(c)}
\end{minipage}
\hspace{.05\textwidth}
\begin{minipage}[t]{.4\textwidth}
\hspace{-.05\textwidth} \scalebox{.84}
{\includegraphics*[width=1.17\textwidth]{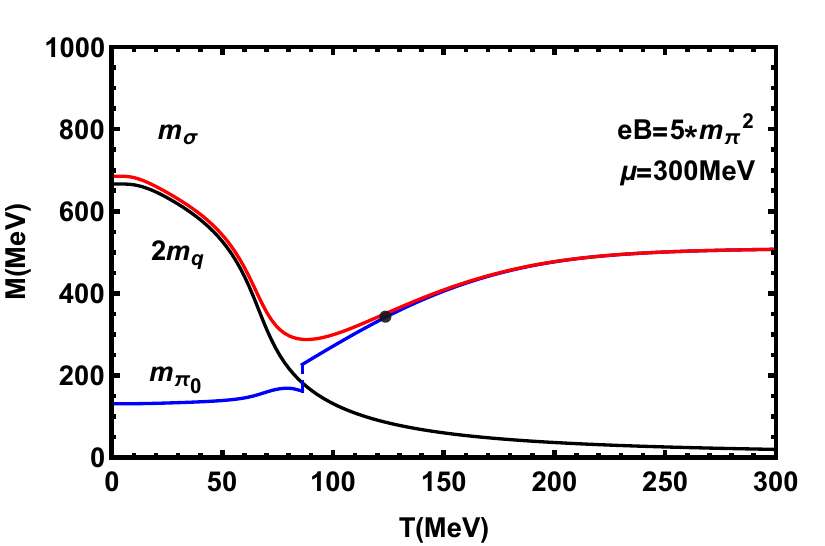}}\\
\centerline{(d)}
\end{minipage}
\parbox{15cm}
{\caption{The masses of quark $\sigma$ and $\pi_0$ mesons at  $\mu=0, 100, 200, 300$ MeV when $eB=5m_{\pi}^{2}$.}
 \label{fig3}}
\end{figure}

      Fig.~\ref{fig2} indicates that with the increase of magnetic field, the initial temperature when the chiral symmetry being restored is increasing continuously, this temperature is so-called critical temperature $T_{C}$. We plot black points in Fig.~\ref{fig2} and Fig.~\ref{fig3}, which means at this point, the chiral symmetry is restored, the temperature of this point is the critical temperature of chiral phase transition $T_{C}$. 

 When $m_{\pi}=2m_{q}$, the temperature is Mott temperature~\cite{Fayazbakhsh:2013cha,Mao:2019avr}, $T_{Mott}$, where the decay $\pi\rightarrow$ qq is possible to happen. When $m_{\sigma}=2m_{\pi}$ the temperature is the dissociation temperature, $T_{diss}$, where the $\sigma\rightarrow\pi\pi$ decay can happen. In the chiral limit, $T_{Mott} =T_{diss} =T_C$. When $T> T_{diss}$, a $\sigma$ meson can decay into two $\pi $ mesons. When $T> T_{Mott}$, a $\pi$ meson can decay into a pair of positive and negative quarks. When the temperature is $T_{Mott}$, there is a distinct transition in $m_{\pi}$. There is also a distinct transition from $m_{\sigma}>2m_{\pi}$ to $m_{\sigma}\approx 2m_{\pi}$ at $T_{Mott}$. So we set $T_{Mott} \approx T_{diss}$ in our work, and for the strong magnetic field this relation is also fulfilled.

The phase transition with temperature is a crossover, and we see that in the crossover region, the mass of the $\sigma$ meson increases with the increase of the magnetic field. However in the crossover region, the mass of $\pi_{0}$ meson decreases slightly with the increase of magnetic field, this result is same with that in Ref.~\cite{Preis:2010cq}. When the chiral symmetry is restored, the mass of $\sigma$ meson is degenerate with that of $\pi_{0}$ meson. At a finite temperature, the mass of the constituent quarks increases with the enhancement of the magnetic field, which is due to the fact that the external magnetic field can also cause the magnetic catalysis (MC)~\cite{Inagaki:2003yi,Klimenko:1991he,Gusynin:1995nb}. In  Fig.~\ref{fig2}, we can clearly see that $T_{C}$ and $T_{Mott}$ are all increasing, and when the magnetic field is strong enough, $T_{Mott} \approx T_{C}$.

Fig.~\ref{fig3} are the diagrams of mass change corresponding to the temperature under different chemical potential $\mu=0, 100, 200, 300$ MeV, where we fixed the magnetic filed $eB=5m_{\pi}^{2}$. From these figures, it is observed that when the temperature increase, the mass of quark decreases slowly and then decreases quickly. Finally the mass of quark changes slowly around $ T = 200, 190, 160, 90$ MeV and tends to stay unchanged in different chemical potential. For a fixed temperature, we find that as the chemical potential increases, the mass of quark gradually decreases.

When the chiral symmetry is restored, the mass of $\sigma$ meson degenerates to the mass of $\pi_0$ meson. Here one can clearly observe that for a fixed strong external magnetic field, with increasing of chemical potential, Mott transition temperature $T_{Mott}$ and the critical temperature $T_{C}$ for the restoration of chiral symmetry are all decreasing gradually. In the region of crossover, the mass of $\sigma$ meson decreases continually with the chemical potential, while the mass of $\pi_0$ increases with the chemical potential. There are still a sharp change in the mass of $\pi_0$ for the existence of the magnetic filed.

In fact, there exists some non-monotonical behavior of pion mass near the critical temperature, which does not conflict with the chiral symmetry restoration process. This behavior was predicted by Son and Stephanov in Ref. \cite{Son:2001ff} by scaling and universality arguments, and also observed later by the lattice results in Ref. \cite{Brandt:2014qqa}. This behavior is due to the interplay between the pion’s velocity and its screening mass at finite temperature. The pion’s velocity drops near $T_C$, while the screening mass increases with temperature.

 \begin{figure}[htb]
\hspace{-.05\textwidth}
\begin{minipage}[t]{.4\textwidth}
\includegraphics*[width=\textwidth]{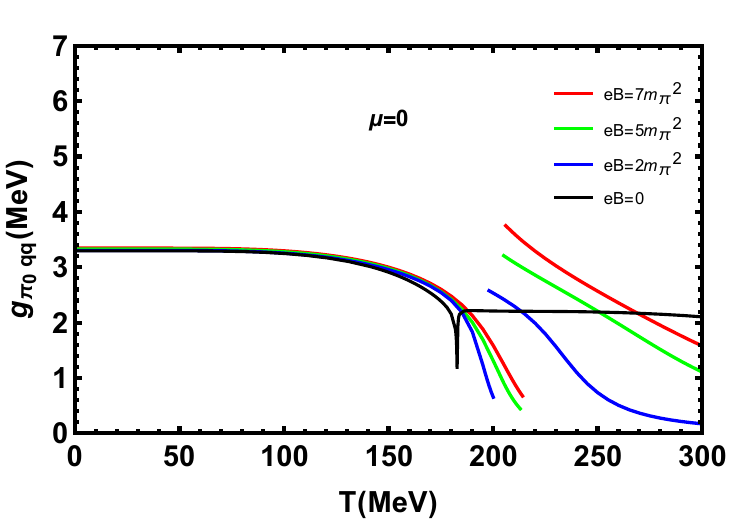}\\
\centerline{(a)}
\end{minipage}
\hspace{.05\textwidth}
\begin{minipage}[t]{.4\textwidth}
\hspace{-.05\textwidth} \scalebox{.84}
{\includegraphics*[width=1.17\textwidth]{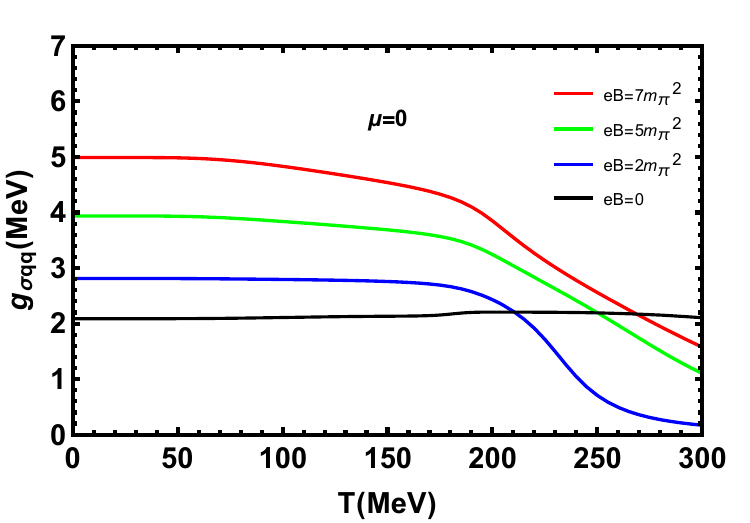}}\\
\centerline{(b)}
\end{minipage}
\parbox{15cm}
{\caption{The coupling constant of $g_{\pi_0 qq}$  and $g_{\sigma qq}$ at different magnetic field when the chemical potential $\mu=0$.}
 \label{fig4}}
\end{figure}

Fig.~\ref{fig4} shows the corresponding coupling constant $ g _ {\pi_0 qq} $ and $ g _ {\sigma qq} $ when we change the temperature. Here we fix the chemical potential $\mu=0$ and the magnetic fields are $ eB =0, 2, 5, 7 m_{\pi}^{2}$. One can see that the coupling constant becomes larger with the increase of the magnetic field. For a fixed magnetic field, the coupling constant is basically constant at low temperature, then it changes significantly around $ T = 200$ MeV.

In Fig.~\ref{fig4}.(a), the corresponding coupling constant $ g _ {\pi_0 qq} $ increases slightly with the increase of magnetic field when we choose a fixed temperature. In Fig.~\ref{fig4}.(b), the corresponding coupling constant $ g _ {\sigma qq} $ also increase with the increase of magnetic field. 
 When it comes to $T_{Mott}$, the $ g _ {\pi_0 qq} $ drops down significantly and then jump to a high value.    
 
  \begin{figure}[htb]
\hspace{-.05\textwidth}
\begin{minipage}[t]{.4\textwidth}
\includegraphics*[width=\textwidth]{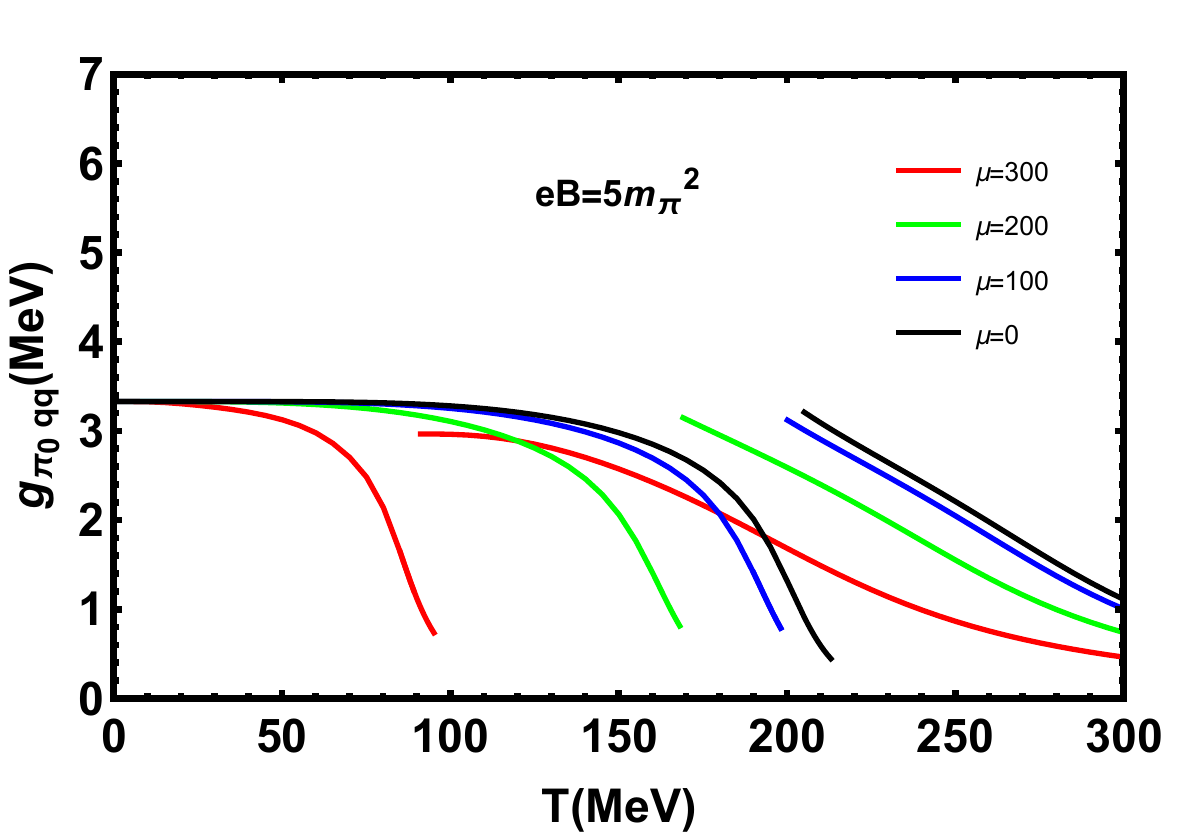}\\
\centerline{(a)}
\end{minipage}
\hspace{.05\textwidth}
\begin{minipage}[t]{.4\textwidth}
\hspace{-.05\textwidth} \scalebox{.84}
{\includegraphics*[width=1.17\textwidth]{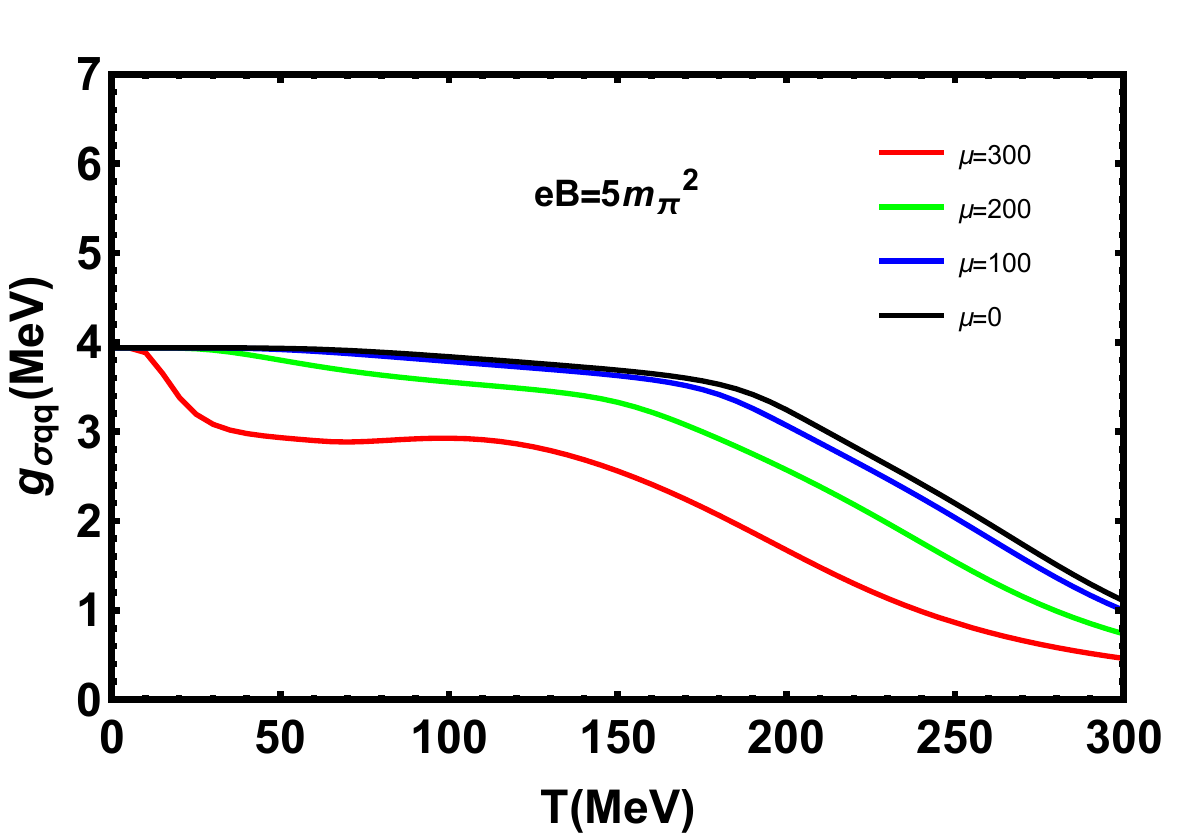}}\\
\centerline{(b)}
\end{minipage}
\parbox{15cm}
{\caption{The coupling constant of $g_{\pi_0 qq}$ and $g_{\sigma qq}$ at different chemical potential when the magnetic field $eB=5m_{\pi}^{2}$.}
 \label{fig5}}
\end{figure}

Fig.~\ref{fig5} shows the effects of the chemical potential through coupling constant $ g _ {\pi_0 qq} $ and $ g _ {\sigma qq} $. Here we fix the magnetic field $eB=5m_{\pi}^{2}$ and the chemical potential are $\mu=0, 100, 200, 300$ MeV. In Fig.~\ref{fig5}.(a), the coupling constant becomes smaller with the increase of the chemical potential. For a fixed magnetic field, the coupling constant is basically constant and then changes significant when the chemical potential is finite. One can clearly see that the critical temperature when $ g _ {\pi_0 qq}$ jumps is decreasing with the increase of chemical potential, which means adding chemical potential suppresses the decay of the $\pi_0$ meson. In Fig.~\ref{fig5}.(b), the chemical potential has a significant impact on coupling constant $ g _ {\sigma qq} $, and it decreases gradually with the increase of temperature. 

\begin{figure}
  \centering
\includegraphics[width=3.12in,height=2.39in]{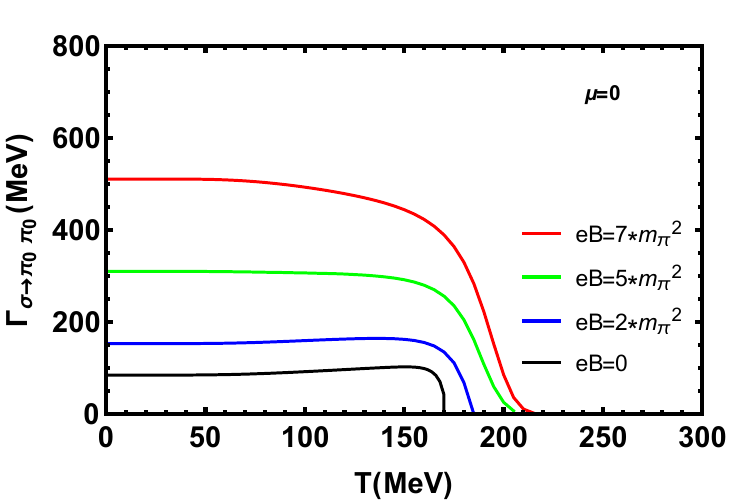}\\
 { \caption{The decay width of $\Gamma _{\sigma\rightarrow\pi_0\pi_0}$ at different magnetic fields /when the chemical potential $\mu=0$.}\label{fig6}}
\end{figure}

\begin{figure}
  \centering
\includegraphics[width=3.12in,height=2.39in]{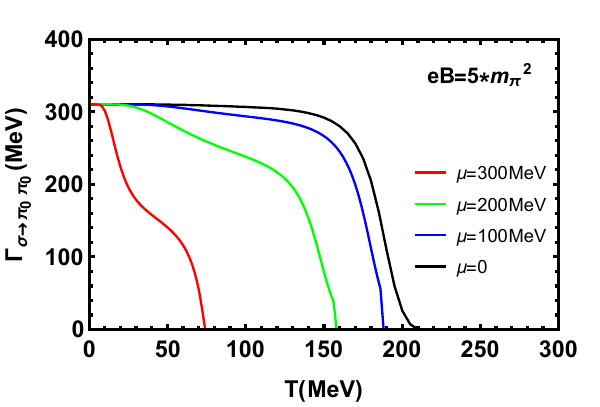}\\
  {\caption{The decay width of $\Gamma _{\sigma\rightarrow\pi_0\pi_0}$ at different chemical potentials when the magnetic field $eB=5m_{\pi}^{2}$.}
\label{fig7}}
\end{figure}

The decay width $\Gamma _ {\sigma\rightarrow\pi_0\pi_0} $ with temperature are shown in Fig.~\ref{fig6} and Fig.~\ref{fig7}. We fix the chemical potential $\mu=0$ in Fig.~\ref{fig6}, and the magnetic field are respectively $ eB =0, 2, 5, 7 m_{\pi}^{2}$. We find that when the magnetic field increase, the decay width increase. In Lattice QCD~\cite{Ding:2020hxw}, the decay width of $\pi_{0}$ have been calculated. The decay constant also shows an increase with the increase of magnetic field. When $ eB = 0$, the decay width is 84.7 MeV, and when the magnetic field is added, the decay width becomes 153.1 MeV, 310.2 MeV and 510.6 MeV respectively. It is about 1.8 times, 3.7 times and 6.0 times of that in the case of zero magnetic field. And with the increase of temperature, the decay critical point also increases from $ T = 170 $ MeV to $ T = 220 $ MeV. Finally the decay width $ \Gamma _ {\sigma\rightarrow\pi_0\pi_0} $ decreases, and goes to 0 when the temperature is around the critical point temperature. At this temperature, the decay happens rapidly and then stops.

In Fig.~\ref{fig7} we fix the magnetic field $eB = 5m _ {\pi} ^ {2}$, and the chemical potential are $ \mu=0, 100, 200, 300$ MeV. One find that when the chemical potential increase, the corresponding decay width $ \Gamma _ {\sigma\rightarrow\pi_0\pi_0} $ decreases, and the decay critical points move to lower temperature, their temperatures are respectively 210, 189, 159, 72 MeV. Nearby the critical point, $ \Gamma _ {\sigma\rightarrow\pi_0\pi_0} $ suddenly changes to 0, which means the decay process stops. It is clear that the presence of the chemical potential suppresses the decay of $\sigma$ meson.

     Under the corresponding magnetic field in Fig.~\ref{fig7}, the decay width increases sharply at $T_{Mott}$ when we add the magnetic field, which reflects the restoration trend of chiral symmetry broken. Similar results on the decay widths of $\sigma$ meson at fixed chemical potential $\mu=0$ can be found in Ref.~\cite{Costa:2004xw} to the study of the decay widths of neutral $\pi$ meson decaying into photons for different neutral mesons. The increase of decay width is particularly pronounced when the magnetic field is very large. When we consider different chemical potential and fix the external magnetic field $eB = 5m _ {\pi} ^ {2}$ in Fig.~\ref{fig7}, the change of $ \Gamma_{\sigma\rightarrow\pi_0\pi_0} $ looks very clear and the critical point decreases with the chemical potential .

 In Ref.~\cite{Fayazbakhsh:2013cha}, they have considered weak decay constant of $\pi_{0}$ at finite temperature T, chemical potential $\mu$ and in the presence of a constant magnetic field B. In their work, they consider the decay constant of $\pi_{0}$ at $\mu=0$. While in our paper, we mainly calculate the decay constant of $\sigma$ with different chemical potential and magnetic field, that is the new point in our paper. We also calculate the mass spectra of mesons. 

    From all the figures, one may find that the critical temperature of $ g _ {\pi_0 qq} $, $ g _ {\sigma qq} $ and $ \Gamma _ {\sigma\rightarrow\pi_0\pi_0} $ is $T_{Mott}$, not $T_{C}$. In Refs.~\cite{Zhuang:1994dw,Zhu:2001iu,Zhuang:2000tz}, the $ \Gamma _ {\sigma\rightarrow\pi\pi} $ is usually associated with chiral phase transition, in their work, the threshold temperature is $T_{diss}$ since there is no magnetic field. With the magnetic field, $T_{C}$ and $T_{Mott}$ are different, this difference is studied in Refs.~\cite{Mao:2016lsr,Mao:2016,Mao:2019avr}. $T_{Mott}$ is decreasing with the increase of chemical potential, which is same with~\cite{Zhu:2001iu}. In this paper, $ \Gamma _ {max} $ is first decreasing and then increasing with the increase of chemical potential, when the chemical potential is bigger enough, it turns to zero. While in our work, the change of $ \Gamma _ {\sigma\rightarrow\pi_0\pi_0} $ is different since we chose a constant magnetic field $eB = 5m _ {\pi} ^ {2}$. With the increase of chemical potential, $ \Gamma _ {\sigma\rightarrow\pi_0\pi_0} $ is decreasing clearly.

\section{Summary and Conclusion}\label{sec:S}
	
    Magnetic field was generated in the early universe. It can influence subsequent cosmic phase transitions, which is important in particle physics in the early universe. Our results suggest that the presence of an external magnetic field "magnetizes" quarks in neutral mesons, affecting their thermodynamic properties. The existence of the magnetic field increases the critical temperature, thus increases the symmetry breaking region and suppresses the phase transition to some extent.
	
	From this work, we clearly see that the neutral meson is also affected by the external magnetic field, which is mainly due to the “magnetization” of the charged quarks that make up the meson, which affects the properties of the meson, such as meson mass, coupling constant, decay width and so on. Here, the effect of the magnetic field is to increase the critical temperature in the mass spectrum, thereby to increase the symmetry breaking region. At the same time, the existence of magnetic field breaks the isospin symmetry, and the separation of quark energy levels leads to the jump in $\pi_0$ mass.
	
	The effect of the chemical potential is to decrease the critical temperature in the mass spectrum. In the case of fixed strong external magnetic field $eB=5m_{\pi}^{2}$, with the increase of chemical potential, the masses of $\sigma$, $\pi_0$ and quark don't change much at low temperature, while the coupling constant and decay width all decrease. The existence of chemical potential reduces the critical temperature and accelerated the restoration of chiral symmetry breaking. It is worth mentioning that the critical temperature of $ g _ {\pi_0 qq} $, $ g _ {\sigma qq} $ and $ \Gamma _ {\sigma\rightarrow\pi_0\pi_0} $ is $T_{Mott}$, not $T_{C}$. We know that the truly critical temperature of $\sigma\rightarrow\pi_0\pi_0$ should be $T_{diss}$, $T_{Mott} \approx T_{diss}$ when we consider the magnetic field in our work.

     To conclude, we use SU(2) NJL model to study the decay process of $\sigma\rightarrow\pi_0\pi_0$, and our results show that the magnetic field will enhance the decay process while the chemical potential will reduce the decay process. In this work, a general method is used to calculate the decay constant, which can be extended to other charged mesons, such as $K$ or $\rho$ mesons. It can also be extended to other background , such as 3-flavor NJL model, PNJL model and so on.

\section*{Acknowledgments}

This work is in part supported by the NSFC Grant Nos. 11735007, 11890711 and 11890710. 

\bibliographystyle{unsrt}
\bibliography{Final}

\begin{thebibliography}{65}

\bibitem{Goldberger:1958tr} 
  M.~L.~Goldberger and S.~B.~Treiman,
  Phys.\ Rev.\  {\bf 110}, 1178 (1958).

\bibitem{GellMann:1968rz} 
  M.~Gell-Mann, R.~J.~Oakes and B.~Renner,
  Phys.\ Rev.\  {\bf 175}, 2195 (1968).

\bibitem{Soloveva:2019xph} 
  O.~Soloveva, P.~Moreau and E.~Bratkovskaya,
  arXiv:1911.08547 [nucl-th].

\bibitem{Fayazbakhsh:2012vr} 
  S.~Fayazbakhsh, S.~Sadeghian and N.~Sadooghi,
  Phys.\ Rev.\ D {\bf 86}, 085042 (2012)
  [arXiv:1206.6051 [hep-ph]].

\bibitem{Fayazbakhsh:2013cha} 
  S.~Fayazbakhsh and N.~Sadooghi,
  Phys.\ Rev.\ D {\bf 88}, no. 6, 065030 (2013)
 [arXiv:1306.2098 [hep-ph]].

\bibitem{Fayazbakhsh:2013em} 
  S.~Fayazbakhsh and N.~Sadooghi,
  PoS Confinement X {\bf }, 294 (2012)
   [arXiv:1302.0622 [hep-ph]].

\bibitem{Kayanikhoo:2019ugo} 
  F.~Kayanikhoo, K.~Naficy and G.~H.~Bordbar,
  arXiv:1911.10512 [nucl-th].

\bibitem{Gong:2019khm} 
  Z.~Gong, F.~Mackenroth, X.~Q.~Yan and A.~V.~Arefiev,
  Sci.\ Rep.\  {\bf 9}, no. 1, 17181 (2019).

\bibitem{Xu:2021}
  K. Xu, J. Chao and M. Huang, 
Phys. Rev. D 103, 076015 (2021) 
[arXiv:2007.13122 [hep-ph]].
 
\bibitem{Xu:2020}
K. Xu, S. Shi, H. Zhang, D. Hou, J. Liao and M. Huang,
Phys. Lett. B {\bf 809} (2020) 135706
[arXiv:2004.05362 [hep-ph]]. 

\bibitem{Andersen:2014xxa} 
  J.~O.~Andersen, W.~R.~Naylor and A.~Tranberg,
  Rev.\ Mod.\ Phys.\  {\bf 88}, 025001 (2016)
  [arXiv:1411.7176 [hep-ph]].

\bibitem{Miransky:2015ava} 
  V.~A.~Miransky and I.~A.~Shovkovy,
  Phys.\ Rept.\  {\bf 576}, 1 (2015)
  [arXiv:1503.00732 [hep-ph]].
 
\bibitem{Li:2016tel}
H.~Li, X.~l.~Sheng and Q.~Wang,
Phys. Rev. C \textbf{94}, no.4, 044903 (2016)
[arXiv:1602.02223 [nucl-th]].

\bibitem{Siddique:2021smf}
I.~Siddique, X.~L.~Sheng and Q.~Wang,
[arXiv:2106.00478 [nucl-th]].

\bibitem{Selyuzhenkov:2005xa} 
  I.~V.~Selyuzhenkov [STAR Collaboration],
  Rom.\ Rep.\ Phys.\  {\bf 58}, 049 (2006)
  [nucl-ex/0510069].

\bibitem{Kharzeev:2007jp} 
  D.~E.~Kharzeev, L.~D.~McLerran and H.~J.~Warringa,
  Nucl.\ Phys.\ A {\bf 803}, 227 (2008)
  [arXiv:0711.0950 [hep-ph]].

\bibitem{Skokov:2009qp} 
  V.~Skokov, A.~Y.~Illarionov and V.~Toneev,
  Int.\ J.\ Mod.\ Phys.\ A {\bf 24}, 5925 (2009)
  [arXiv:0907.1396 [nucl-th]].

\bibitem{Ferrer:2010wz} 
  E.~J.~Ferrer, V.~de la Incera, J.~P.~Keith, I.~Portillo and P.~L.~Springsteen,
  Phys.\ Rev.\ C {\bf 82}, 065802 (2010)
  [arXiv:1009.3521 [hep-ph]].

\bibitem{Ayala:2002qy} 
  A.~Ayala, P.~Amore and A.~Aranda,
  Phys.\ Rev.\ C {\bf 66}, 045205 (2002)
  [hep-ph/0207081].

\bibitem{Nambu:1961fr} 
  Y.~Nambu and G.~Jona-Lasinio,
  Phys.\ Rev.\  {\bf 124}, 246 (1961).

\bibitem{Klevansky:1992qe} 
  S.~P.~Klevansky,
  Rev.\ Mod.\ Phys.\  {\bf 64}, 649 (1992).

\bibitem{Hatsuda:1994pi} 
  T.~Hatsuda and T.~Kunihiro,
  Phys.\ Rept.\  {\bf 247}, 221 (1994)
  [hep-ph/9401310].

\bibitem{Buballa:2003qv} 
  M.~Buballa,
  Phys.\ Rept.\  {\bf 407}, 205 (2005)
  [hep-ph/0402234].

\bibitem{kapusta_gale_2006}
Joseph~I. Kapusta and Charles Gale.
\newblock {\em Finite-Temperature Field Theory: Principles and Applications}.
\newblock Cambridge Monographs on Mathematical Physics. Cambridge University
  Press, 2 edition, 2006.

\bibitem{KUNIHIRO1988385}
Teiji Kunihiro and Tetsuo Hatsuda.
\newblock {\em Physics Letters B}, 206(3):385 -- 390, 1988.

\bibitem{Nambu:1961tp}
Yoichiro Nambu and G.~Jona-Lasinio.
\newblock {\em Phys. Rev.}, 122:345--358, 1961.

\bibitem{Zhuang:1994dw} 
  P.~Zhuang, J.~Hufner and S.~P.~Klevansky,
  Nucl.\ Phys.\ A {\bf 576}, 525 (1994).

\bibitem{Ritus:1972ky} 
  V.~I.~Ritus,
  Annals Phys.\  {\bf 69}, 555 (1972).

\bibitem{Ritus:1978cj} 
  V.~I.~Ritus,
  Sov.\ Phys.\ JETP {\bf 48}, 788 (1978)
  [Zh.\ Eksp.\ Teor.\ Fiz.\  {\bf 75}, 1560 (1978)].

\bibitem{Leung:2005yq} 
  C.~N.~Leung and S.~Y.~Wang,
  Nucl.\ Phys.\ B {\bf 747}, 266 (2006)
  [hep-ph/0510066].

\bibitem{Elizalde:2000vz} 
  E.~Elizalde, E.~J.~Ferrer and V.~de la Incera,
  Annals Phys.\  {\bf 295}, 33 (2002)
 [hep-ph/0007033].

\bibitem{Menezes:2008qt} 
  D.~P.~Menezes, M.~Benghi Pinto, S.~S.~Avancini, A.~Perez Martinez and C.~Providencia,
  Phys.\ Rev.\ C {\bf 79}, 035807 (2009)
  [arXiv:0811.3361 [nucl-th]].

\bibitem{Fukushima:2009ft} 
  K.~Fukushima, D.~E.~Kharzeev and H.~J.~Warringa,
  Nucl.\ Phys.\ A {\bf 836}, 311 (2010)
  [arXiv:0912.2961 [hep-ph]].
 
\bibitem{Mao:2016lsr} 
  S.~Mao,
  Phys.\ Rev.\ D {\bf 94}, no. 3, 036007 (2016)
   [arXiv:1605.04526 [hep-th]].

\bibitem{Mao:2016} 
  S.~Mao,
Phys.Rev.D 96 (2017) 3, 034004.

\bibitem{Mao:2019avr} 
  S.~Mao,
  arXiv:1908.02851 [nucl-th].

\bibitem{Mao:2018dqe} 
  S.~Mao,
  Phys.\ Rev.\ D {\bf 99}, no. 5, 056005 (2019)
  [arXiv:1808.10242 [nucl-th]].

\bibitem{Zhu:2001iu}
X. ~Zhu and P. ~Zhuang,
Commun. Theor. Phys. \textbf{37}, 431-434 (2002)
doi:10.1088/0253-6102/37/4/431
[arXiv:nucl-th/0110039 [nucl-th]].

\bibitem{Zhuang:2000tz}
P.~Zhuang and Z.~Yang,
Chin. Phys. Lett. \textbf{18}, 344-346 (2001)
[arXiv:nucl-th/0008041 [nucl-th]].

\bibitem{Gatto:2012sp} 
  R.~Gatto and M.~Ruggieri,
  Lect.\ Notes Phys.\  {\bf 871}, 87 (2013)
  [arXiv:1207.3190 [hep-ph]].

\bibitem{Preis:2012fh} 
  F.~Preis, A.~Rebhan and A.~Schmitt,
  Lect.\ Notes Phys.\  {\bf 871}, 51 (2013)
  [arXiv:1208.0536 [hep-ph]].

\bibitem{Mao:2016fha}
S.~Mao,
Phys. Lett. B \textbf{758}, 195-199 (2016)
[arXiv:1602.06503 [hep-ph]].

\bibitem{Preis:2010cq} 
  F.~Preis, A.~Rebhan and A.~Schmitt,
  JHEP {\bf 1103}, 033 (2011)
 [arXiv:1012.4785 [hep-th]].

\bibitem{Inagaki:2003yi} 
  T.~Inagaki, D.~Kimura and T.~Murata,
  Prog.\ Theor.\ Phys.\  {\bf 111}, 371 (2004)
  [hep-ph/0312005].


\bibitem{Klimenko:1991he} 
  K.~G.~Klimenko,
  Z.\ Phys.\ C {\bf 54}, 323 (1992).

\bibitem{Gusynin:1995nb} 
  V.~P.~Gusynin, V.~A.~Miransky and I.~A.~Shovkovy,
  Nucl.\ Phys.\ B {\bf 462}, 249 (1996)
  [hep-ph/9509320].

\bibitem{Son:2001ff}
D.~T.~Son and M.~A.~Stephanov,
Phys. Rev. Lett. \textbf{88}, 202302 (2002)
[arXiv:hep-ph/0111100 [hep-ph]].

\bibitem{Brandt:2014qqa}
B.~B.~Brandt, A.~Francis, H.~B.~Meyer and D.~Robaina,
Phys. Rev. D \textbf{90}, no.5, 054509 (2014)
[arXiv:1406.5602 [hep-lat]].


\bibitem{Ding:2020hxw}
H.~T.~Ding, S.~T.~Li, A.~Tomiya, X.~D.~Wang and Y.~Zhang,
Phys. Rev. D \textbf{104}, no.1, 014505 (2021)
[arXiv:2008.00493 [hep-lat]].


\bibitem{Costa:2004xw} 
  P.~Costa, M.~C.~Ruivo and Y.~L.~Kalinovsky,
  Phys.\ Rev.\ C {\bf 70}, 048202 (2004)
 [hep-ph/0403263].





\end{thebibliography}

\end{document}